\documentclass[pre,twocolumn,superscriptaddress,amsmath,reprint]{revtex4-1}
\usepackage{graphicx}
\usepackage{balance}
\usepackage{color}
\usepackage{balance}
\usepackage{bm}
\usepackage{ulem} 

\renewcommand{\Re}{\mathop{\mathrm{Re}}}
 
\newcommand{\bs}{\mathbf {s}}

\newcommand{\bS}{\mathbf {S}}

\newcommand{\cH}{\mathcal {H}}

\begin{document}

\title{Finite size bath in qubit thermodynamics}

\author{J. P. Pekola}
\affiliation{Low Temperature Laboratory, Department of Applied Physics, Aalto University School of Science, P.O. Box 13500, 00076 Aalto, Finland}
\author{S. Suomela}
\affiliation{Department of Applied Physics and COMP Center of Excellence, Aalto University School
of Science, P.O. Box 11100, 00076 Aalto, Finland}
\author{Y. M. Galperin}
\affiliation{Physics Department, University of Oslo, P.O.Box 1048 Blindern, 0316 Oslo, Norway}
\affiliation{A. F. Ioffe Physico-Technical Institute RAS, 194021 St.
Petersburg, Russian Federation}

\begin{abstract}
We discuss a qubit  weakly coupled to a finite-size heat bath (calorimeter) from the point of view of quantum thermodynamics. The energy deposited to this environment together with the state of the qubit provides a basis to analyze the heat and work statistics of this closed combined system. We present results on two representative models, where the bath is composed of two-level systems or harmonic oscillators, respectively. Finally, we derive results for an open quantum system composed of the above qubit plus finite-size bath, but now the latter is coupled to a practically infinite bath of the same nature of oscillators or two-level systems.
\end{abstract}

\date{\today}

\maketitle
\section{Introduction}
Currently there is considerable interest to understand thermodynamics of small systems,  see, e.g.,~\cite{Jarzynski15} and references therein.
In the classical regime, experiments have been devised that accurately confirm many of the modern relations relating to heat and work statistics in stochastic thermodynamics. Yet in quantum systems, measuring such quantities poses naturally new questions according to general principles of quantum mechanics. One of the measurement strategies in this respect is to observe the heat deposited to classical environment by a thermometer (calorimeter) operating at sub-kelvin temperatures~\cite{Pekola13}. In practice, measuring the full environment, for instance in form of the heat bath, is by no means a trivial task. An ideal bath is infinite, and measuring all of it is not possible. To measure tiny energy exchanges in mesoscopic devices one definitely needs a small-sized calorimeter playing role of a thermal bath for a quantum device. On the other hand, if the bath is finite, which is the topic of the current work, it is not in full equilibrium under the influence of energy (heat) exchanged between the quantum system under study and this mini-bath.
 
It is the aim of this work to present \textit{minimal} theoretical models of finite-size environments and to investigate non-equilibrium behavior and fluctuation relations (FRs) in these set-ups. Within these models, the quantum device is coupled to a finite-size calorimeter, which, in turn, may be coupled with the true bath.
In particular, we will consider to what extent measurement with a finite-size calorimeter can reveal the predictions of the Crooks fluctuation theorem~\cite{Crooks99} and Jarzynski equality~\cite{Jarzynski97a,Jarzynski97b}. We will show that these relationships are satisfied even if the quantum device and the calorimeter form an open system when the initial state is taken to be a factorized canonical configuration. If the calorimeter is coupled to the true bath, then to keep the FRs valid one has to keep track of the energy exchange between the calorimeter and the bath. However, if \textit{all} the energy flow from the device to the calorimeter is measured  at the level of single quanta one can still satisfy the FRs. This is because within our model all the heat flow from the device to the true bath takes place through the calorimeter.

The paper is organized as follows. In Sec.~\ref{description} we describe the  model set-ups to be studied and formulate main equations. Dynamics of a qubit coupled to a finite-size calorimeter is analyzed in Sec.~\ref{dynamics}. Results of numerical analysis of heat and work distributions and their discussion are given in Sec.~\ref{distributions}. In Appendix we present analytic treatment of the FRs for the case when at most a single quantum jump takes place.

\section{Model set-ups and main equations} \label{description}

\begin{figure}[h!]
    \begin{center}
    \includegraphics[width=.9\columnwidth]{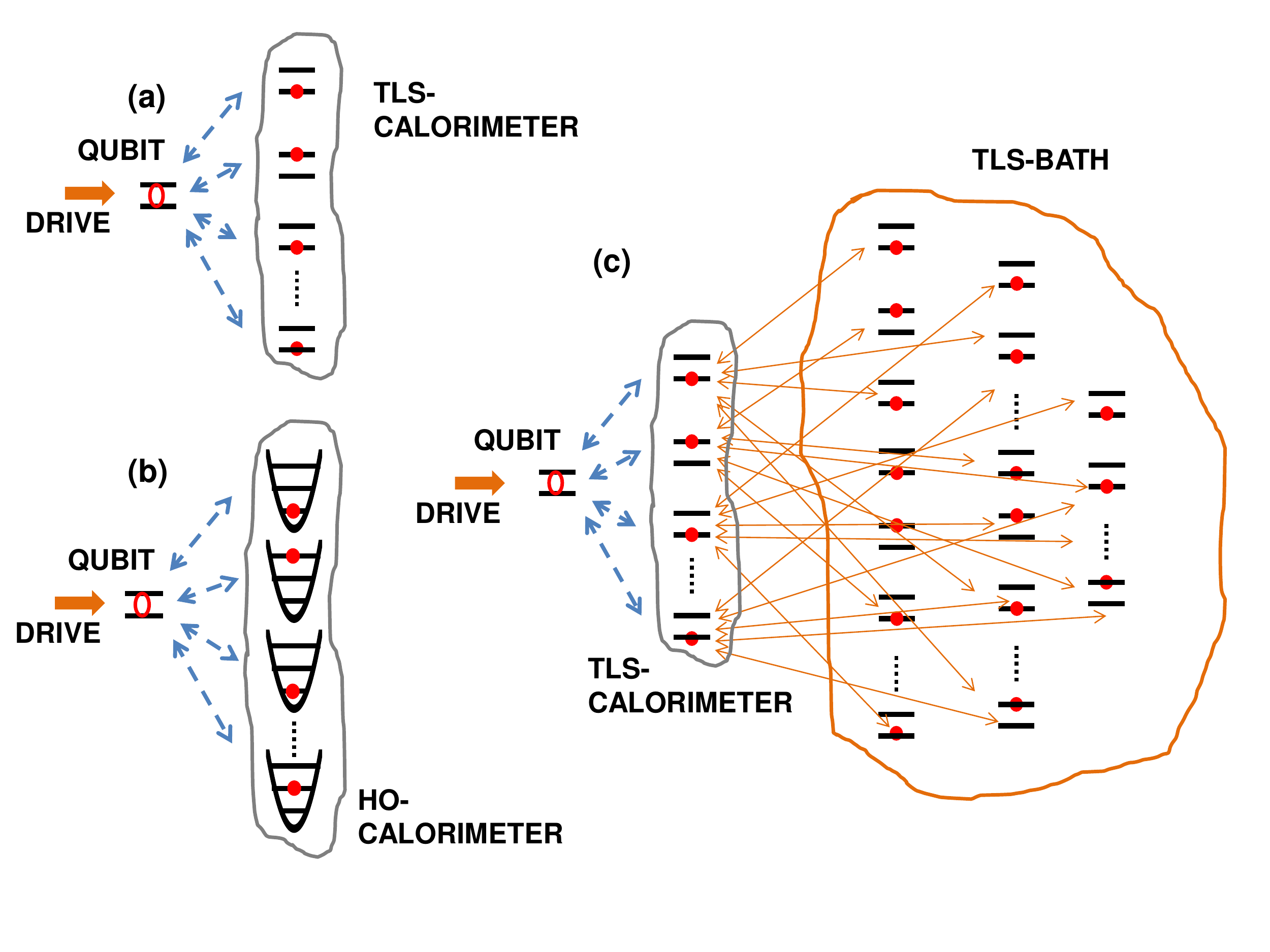}
    \end{center}
    \caption{\label{fig1} Systems considered. (a) Closed system composed of a driven qubit coupled to a calorimeter modelled as a collection of $N_{\text{c}}$ two-level systems (TLSs). (b) Closed system as in (a) but with a harmonic oscillator (HO) based calorimeter. (c) An open system where the calorimeter of (a) is coupled to a large bath of $N_B$ TLSs.}
\end{figure} 
Figure~\ref{fig1} presents schematically the set-ups that we consider here.
They compose of a qubit with energy gap $\hbar\omega_0$, which can be driven by an external classical source. The qubit is coupled to a finite bath of $N_{\text{c}}$ two-level systems (TLSs, Fig.~\ref{fig1}~(a)) or harmonic oscillators (HOs, Fig.~\ref{fig1}~(b)), generically called ``fluctuators", with level spacing of the $k$-th oscillator equal to $\hbar\Omega_k$. We call this bath a ``calorimeter" in what follows. In Fig.~\ref{fig1}~(c) the calorimeter is further coupled to a large bath of $N_{\text{b}}$ fluctuators, mimicking a virtually infinite-size thermal bath. We consider here the case when both the calorimeter and the bath consist of TLSs.
Two-level tunneling systems are generic for disordered systems~\cite{Anderson72,Phillips72}, they couple to quantum devices leading to their decoherence, see~\cite{Paladino14,Bergli09} and references therein for a review.  

We will characterize the qubit by the 1/2-spin
$\bS$ interacting with  fluctuators (two-level systems) of the calorimeter characterized by the 1/2-spins $\bs_i$. Then the Hamiltonian takes the form
\begin{equation}
  \label{eq:001}
  \cH=\cH_{\text{q}}+ \cH_{\text{man}} +\cH_{\text{c}} +\cH_{\text{b}} +\cH_{\text{q-c}} +\cH_{\text{c-b}} \,. 
\end{equation}
Here the qubit Hamiltonian, $\cH_{\text{q}}$,  can be expressed as
\begin{equation}
 \cH_{\text{q}} = \frac{\Delta}{2}\tilde{S}_z+ \frac{\Lambda}{2}\tilde{S}_x = \frac{\hbar \omega_0}{2}S_z, \ \hbar \omega_0=\sqrt{\Delta^2+\Lambda^2}.
\end{equation}
In this expression and in the following ones  $\tilde{S}_i$ denote the qubit pseudo-spin operators in the ``left-right"  
representation while $S_i$ are the operators in the eigen representation of the qubit. 
The qubit is subjected to a ``manipulation" field $F$, described by the Hamiltonian 
\begin{equation} \label{man}
\cH_{\text{man}}=S_x  F \sin \omega t\,.
\end{equation}
The two-level fluctuators in the calorimeter part are specified in a similar way,
\begin{equation} \label{hc}
 \cH_{\text{c}} = \frac{1}{2}\sum_{i \in \text{c}}(\delta_i\tilde{s}_{z,i}+ \lambda_i\tilde{s}_{x,i} )= \frac{1}{2}\sum_{i \in \text{c}} \varepsilon_i s_{z,i},  
\end{equation}
where $\ \varepsilon_i=\sqrt{\delta^2_i+\lambda^2_i}$.
Since both the qubit and the fluctuators interact with a bath we have to add the Hamiltonian of the bath, $\cH_{\text{b}}$.  It is given by the expression similar to Eq.~\eqref{hc}, but with $i\in \text{b}$.
The interaction terms can be specified as 
\begin{eqnarray}
\cH_{\text{q-c}} &=& \frac{1}{2}\sum_i ( v_i S_z s_{z,i}+
u_i  S_x s_{x,i}) \, ,  \\ 
\cH_{\text{c-b}}&=&\frac{1}{2} \sum_{j\in \text{c},k \in \text{b}} u_{jk} s_{x,j}s_{x,k} \, .
\end{eqnarray}
In the above expressions,
$\hbar \omega_0$ is the distance between the qubit
levels, $\varepsilon$ is the distance 
between the fluctuator's levels, $S_i$ and $s_i$ are the Pauli
matrices acting respectively in the spaces of the qubit and the
fluctuator, and $v_i$ and $u_i$ are the diagonal and off-diagonal coupling constants, respectively. The diagonal qubit-fluctuator interaction $\propto S_z s_z$ is responsible for the dephasing of the qubit due to transitions of the fluctuators between their states.  This contribution is mostly important  far from the resonance between the qubit and a fluctuator. A similar Hamiltonian has been derived and investigated in connection with spectral diffusion in glasses~\cite{Joffrin,Black,Laikhtman}. 
On the contrary, the off-diagonal part is important only when the qubit and the fluctuator have nearly the same energy splitting, $|\hbar \omega_0-\varepsilon_i | \ll \hbar \omega_0$. Importance of this	interaction was stressed in Ref.~\onlinecite{Martinis}.
Its manifestations were then studied in several papers, see, 
e.g., \cite{Galperin05,Lisenfeld10,Grabovskij12,Lisenfeld14} and references therein.

For simplicity, we omit direct interaction of the qubit with 
phonons/photons. This interaction is assumed to be included into the energy relaxation rate, $\hbar/\tau_1$, and the dephasing rate, $\hbar/\tau_2$, of the qubit in the Bloch-Redfield equations~\cite{Bloch,Redfield} for the qubit density matrix. The same assumption is made for the TLSs forming the calorimeter and the bath.

We will also assume that
\begin{equation}
  \label{eq:003}
|u| \ll \hbar / \tau_2 \ll \beta^{-1} \ll \hbar \omega_0
\end{equation}
where $\beta=1/(k_B T)$ is the inverse of bath temperature $T$. Under these assumptions one can ignore coherent coupling between the qubit and fluctuators (as well as between different fluctuatotrs) and consider the action of the fluctuators as a stochastic noise. Following the procedure outlined in Ref.~\onlinecite{Shnirman05} we can present the noise spectrum as
\begin{eqnarray}
S(\omega) &=& \sum_i \gamma_i^2\left ( \cos^2 \theta_i (1-\langle s_{z,i}\rangle^2)\frac{2\Gamma_{1,i}}{\Gamma_{1,i}^2 + \omega^2} \right. \nonumber \\
&&+ \left. \sin^2 \theta_i \frac{1+\langle s_{z,i}\rangle}{2}\frac{2\Gamma_{2,i}}{\Gamma_{2,i}^2 + (\omega- \varepsilon_i/\hbar)^2} \right. \nonumber \\
&&+\left.\sin^2 \theta_i \frac{1-\langle s_{z,i}\rangle}{2}\frac{2\Gamma_{2,i}}{\Gamma_{2,i}^2 + (\omega+ \varepsilon_i/\hbar)^2} \right) \, . \label{noise1}
\end{eqnarray}
Here $\gamma_i^2=u_i^2+v_i^2$,  $\tanh \theta_i \equiv u_i /v_i$. 
Note that $\theta_i$ in general depend on the operating points of both the qubit and $i$-th fluctuator.
The  quantity $\langle \tilde{s}_{z,i} \rangle$ is the difference between the populations of the upper and the lower level of the $i$-th fluctuator, 
$\langle \tilde{s}_{z,i} \rangle =n_{i\downarrow}-n_{i\uparrow}$.
 In thermal equilibrium $\langle \tilde{s}_{z,i}\rangle =\tanh(\beta\varepsilon_i/2)$. The quantities $\Gamma_{1,i} \equiv 1/\tau_{1,i}$ and  $\Gamma_{2,i} \equiv 1/\tau_{2,i}$ are, respectively,  the energy relaxation rate and the dephasing rate of the $i$-th fluctuator. Both quantities are assumed to be much smaller than the temperature of the bath, $\beta^{-1}$.

The expression \eqref{noise1} is based on the standard assumption that the calorimeter contains a very large number of TLSs, the expressions $ P_{g,e}^{(i)} =(1\pm \langle s_{z,i}\rangle)/2 $ are just the probabilities for the ground/excited state to be occupied. In a finite calorimeter, one has to take into account instantaneous rather than the average occupation numbers.
Now we have mapped our system on a Bloch-Redfield dynamics of a qubit subjected to a noise induced by the set of dynamic fluctuators forming the ``calorimeter". This noise contains low-frequency components [1st item in Eq.~\eqref{noise1}] and high-frequency components [2nd and 3d items in Eq.~\eqref{noise1}] allowing for the energy transfer. 

\section{Qubit dynamics} \label{dynamics}

We proceed along the standard way to analyze the qubit dynamics. We introduce the reduced density matrix $\sigma = \mathrm{Tr} _{\text{c}}\,  (\rho)$, where $\rho =\sigma \otimes \rho_{\text{c}}$ is the full density matrix, and the trace $\mathrm{Tr}_{\text{c}}$ is taken over the degrees of freedom of the calorimeter. 

The set of equations for the elements of the qubit density matrix can be cast in the form
\begin{widetext}
\begin{eqnarray} \label{m21}
\dot \sigma_{gg}(M_{\text{c}}) &=& -\frac{F}{\hbar}\Re [\sigma_{ge}(M_{\text{c}}) e^{i(\omega_0-\omega )t}]-\Gamma_\uparrow(M_{\text{c}}) \sigma_{gg}(M_{\text{c}}) + \Gamma_\downarrow (M_{\text{c}}+1) \sigma_{gg} (M_{\text{c}}+1),
\nonumber \\
\dot \sigma_{ee}(M_{\text{c}}) &= &\phantom{-}\frac{F}{\hbar}\Re [\sigma_{ge}(M_{\text{c}}) e^{i(\omega_0-\omega )t}]-\Gamma_\downarrow(M_{\text{c}}) \sigma_{ee}(M_{\text{c}}) + \Gamma_\uparrow (M_{\text{c}}-1) \sigma_{ee} (M_{\text{c}}-1),
\nonumber \\
\dot\sigma_{ge}(M_{\text{c}}) &=&\phantom{-} \frac{F}{2\hbar}e^{i(\omega-\omega_0)t}[\sigma_{gg}(M_{\text{c}})-\sigma_{ee}(M_{\text{c}})]-\frac{1}{2}[\Gamma_\uparrow (M_{\text{c}}) +\Gamma_\downarrow (M_{\text{c}})]\sigma_{ge}(M_{\text{c}}),
\end{eqnarray}
\end{widetext}
where $g$ and $e$ subscripts refer to the ground state and the excited state of the qubit. The first terms in the right-hand sides can be easily obtained by direct commuting of the density matrix with the $\cH_{\text{man}}$ (in the interaction representation). The excitation, $\Gamma_\uparrow$, and relaxation, $\Gamma_\downarrow$, rates are then given by
\begin{equation} \label{m22}
\Gamma_{\uparrow,\downarrow} = (2\pi/\hbar) \sum_i \gamma_i^2 \sin^2 \theta_i 
P^{(i)}_{e,g} \, 
\delta(\hbar \omega_0-\varepsilon_i)\, .
\end{equation}
They can be calculated using representation of the spin operators $\bS$ and $\bs$ through pseudo-fermions, as it was done for the cases of a spin interacting with electrons~\cite{Abrikosov67a,Abrikosov67b} or a TLS interacting with phonons \cite{Maleev80,Maleev83}, or for $S_zs_z$ interaction between two TLSs~\cite{Galperin84}. While deriving Eq.~\eqref{m21} we employed the rotating wave approximation~\cite{Slichter}.

In a finite calorimeter,  the rates depend on the instantaneous state of the calorimeter through the factors $P^{(i)}_{e}=\delta_{n_i, e}$ or $P^{(i)}_{g}=\delta_{n_i,g}$ where $\delta_{a,b}$ is the Kronecker symbol.
Then we can express the ``up" and ``down" rates through the total number of the calorimeter TLSs,
\begin{equation}
N_{\text{c}}=\nu_0 \, \mathcal{V}_{\text{c}} ( \hbar/\tau_2)\, .
\end{equation}
Here $\nu_0$ is the density of the TLSs' states assumed to be constant, and $\mathcal{V}_{\text{c}}$ is the volume of the calorimeter.
Assuming that $M_{\text{c}}$ of the calorimeter TLSs (the number determined by the incoherent dynamics of the calorimeter fluctuators interacting with the qubit and below with the bath) are in the ground state and that all the TLSs are coupled to the qubit with equal coupling constant $\gamma$, $\gamma_i |\sin (\theta_i)| \equiv \gamma$, we can express the rates as
 \begin{equation} \label{m25}
\Gamma_\downarrow = g_{\text{c}}^2 (M_{\text{c}}/N_{\text{c}}), \quad \Gamma_\uparrow = g_{\text{c}}^2 (1-M_{\text{c}}/N_{\text{c}}).
\end{equation}
The effective coupling is then given by $g_{\text{c}}^2 \equiv 2\pi \gamma^2\,\nu_0 \mathcal{V}_{\text{c}}/\hbar $. In this way we represent the calorimeter by $N_{\text{c}}$ TLSs that resonantly interact with the qubit and relax to a bath. Taking into account only these TLSs we can assign a heat capacity to the calorimeter of TLSs as
\begin{equation} \label{hc16}
C=\frac{\partial\langle E\rangle}{\partial T} = N_{\text{c}} k_B \frac{(\beta \hbar\omega_0)^2 e^{\beta\hbar\omega_0}}{(1+e^{\beta \hbar\omega_0})^2},
\end{equation}
where $E$ is the total energy of the  TLSs.
The calorimeter we discuss in the numerical examples below has thus a heat capacity $C/k_B \approx 2$, which is  smaller than what one might obtain experimentally in a typical set up at sub-kelvin temperatures ($C/k_B \approx 10^2...10^3$)~\cite{Pekola13}. The effects that we discuss are thus enhanced correspondingly beyond those expected in the experiment.

We can ascribe to the calorimeter an effective temperature, $T_{\text{c}}$, assuming that the relative number of resonant TLSs  in the ground state in the calorimeter, $ M_{\text{c}}/N_{\text{c}}$, is
\begin{equation} \label{eft}
M_{\text{c}}/N_{\text{c}}=\left( 1+e^{-\beta_{\text{c}}\hbar \omega_0}\right)^{-1}.
\end{equation}
Then
\begin{equation} \label{tc1}
k_B T_{\text{c}}=\beta_{\text {c}}^{-1}=\hbar \omega_0 \ln^{-1} \left(
\frac{M_{\text{c}}}{N_{\text{c}}- M_{\text{c}}} \right).
\end{equation}
Note that $M_{\text{c}}$ is a stochastic variable, and therefore $T_{\text{c}}$ is also a stochastic one. Its distribution function is related to the distribution of $M_{\text{c}}$, $P(M_{\text{c}})$, as
\begin{equation} \label{dft}
P(T_{\text{c}})=\sum_{M_{\text{c}}} P(M_{\text{c}})\delta\left(T_{\text{c}}-\frac{\hbar \omega_0}{k_B \ln \left(\frac
{M_{\text{c}}}{N_{\text{c}}- M_{\text{c}}} \right)} \right).
\end{equation}

If the calorimeter would be composed of $N_{\text{c}}$ harmonic oscillators (HOs) instead, see Fig.~\ref{fig1} (b), we would obtain with a similar procedure in place of Eqs.~\eqref{m25}, cf. with~\cite{Maleev83}
\begin{equation} \label{m26}
\Gamma_\downarrow =\tilde {g}_{\text{c}}^2\sum_{k=1}^{N_{\text{c}}} (1+\mathcal{N}_k),
\quad \Gamma_\uparrow =  \tilde{g}_{\text{c}}^2\sum_{k=1}^{N_{\text{c}}} \mathcal{N}_k.
\end{equation}
Here $\mathcal{N}_k$
refers to the state of the $k$-th HO nearly degenerate with the qubit, and $\tilde {g}_{\text{c}}^2$ is the effective coupling between the qubit and a HO. This result is consistent with that in 
Ref.~\onlinecite{suomela2016} with appropriate definition of the energy of the ensemble of HOs. 

\subsection{Open system with TLS calorimeter and bath}

In the following we consider the case when the finite TLS  calorimeter is further coupled to another bath, see Fig.~\ref{fig1} (c). This is a generalization of the model considered in
\cite{suomela2016}. We assume that the latter is composed of $N_{\text{b}}$ TLSs, with the same level spacing as the former one, and the qubit is not directly interacting with this bath but via the calorimeter only. Typically we consider the case $N_{\text{b}} \gg N_{\text{c}}$, i.e., we take the bath to be virtually infinite. In this bath, like in the calorimeter, all the TLSs are in their eigenstates, $M_{\text{b}}$ of them in the ground state, and the remaining $N_{\text{b}}-M_{\text{b}}$ in the excited state. 
With the same approach as in the earlier sections, one finds the rates in the calorimeter due to the coupling to the bath as
$$ 
\Gamma_{\downarrow,\text{c}} = g^2_{\text{b}} \left(1- \frac{M_{\text{c}}}{N_{\text{c}}}\right)\frac{M_{\text{b}}}{N_{\text{b}}}, 
\  \Gamma_{\uparrow,\text{c}} = g_{\text{b}}^2 \frac{M_{\text{c}}}{N_{\text{c}}}\left(1-\frac{M_{\text{b}}}{N_{\text{b}}}\right).
$$ 
Here $g_{\text{b}}^2$ is the coupling between the calorimeter and the bath, in analogous way to $g_{\text{c}}^2$ of the system-calorimeter coupling but now further normalized by the number of TLSs in the calorimeter. Assuming $N_{\text{b}} \gg 1$ and introducing the bath temperature $\beta_{\text{b}}^{-1}$ we get:
\begin{eqnarray} \label{m25bb}
\Gamma_{\downarrow,\text{c}} & =& g^2_{\text{b}} \left(1- \frac{M_{\text{c}}}{N_{\text{c}}}\right)
\frac{1}{1+e^{-\beta \hbar \omega_0}}\, ,
\nonumber \\
  \Gamma_{\uparrow,\text{c}} &=& g_{\text{b}}^2 \frac{M_{\text{c}}}{N_{\text{c}}} \frac{1}{1+e^{\beta \hbar \omega_0}}\, .
\end{eqnarray}
\begin{figure}[b]
    \begin{center}
    \includegraphics[width=.98\columnwidth]{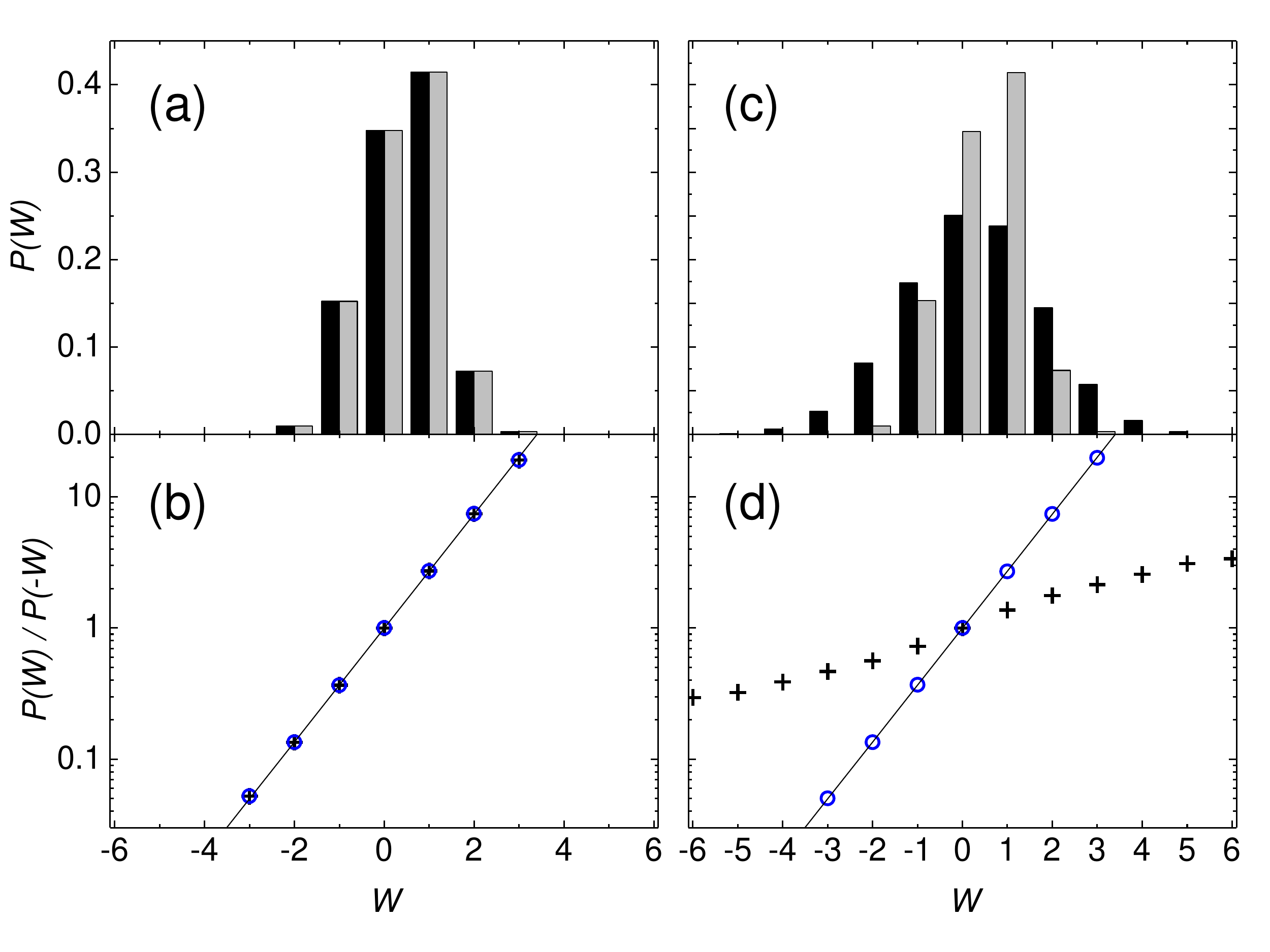}
    \end{center}
    \caption{\label{fig2} Results of a numerical simulation ($10^6$ repetitions) on a system of Fig.~\ref{fig1} (c). The parameters are $N_{\text{c}}=10$, $N_{\text{b}}=10^4$, $\beta\hbar\omega_0=1$, $g_{\text{c}}^2 =0.1$, $F/2\hbar\omega_0=0.05$. The qubit is driven by a $\pi$-pulse. The measurement of the heat is done at the end of the $\pi$-pulse. $U$ is the change of the internal energy of the qubit. In (a) the distribution of $W=U+Q_{\text{c}}$ (black) and $W=U+Q_{\text{b}}+Q_{\text{c}}$ (grey) are shown as a function of the corresponding apparent work $W$. All these quantities are normalized by the level spacing $\hbar\omega_0$. Here $g_{\text{b}}=0$. (b) Test of the Crooks relation (indicated by the straight solid line) for the two expressions of work, $U+Q_{\text{c}}$ (black crosses) and $W=U+Q_{\text{b}}+Q_{\text{c}}$ (blue circles). In (a) and (b) the two sets of data coincide, since the qubit plus calorimeter form a closed system. (c) and (d) show the data as in (a) and (b), respectively, but now for $g_{\text{b}}^2=0.1$. In this case Crooks relation is valid only for $W=U+Q_{\text{b}}+Q_{\text{c}}$.}
\end{figure} 
\subsection{Dynamics of the open system}
When we consider the open system formed of the qubit  driven by $\cH_{\text{man}}$ \eqref{man}
the evolution of the system can be modeled as follows based on the quantum trajectories~\cite{Dalibard92}. We write the stochastic wavefunction of the qubit as
\begin{equation} \label{m25b}
|\psi(t)\rangle =c (t)|g\rangle +d(t)|e\rangle.
\end{equation}
Take a short time interval $\Delta t$, during which at most one jump, either between the qubit and the calorimeter, or between the calorimeter and the bath, can occur. Assume that prior to this time step, $M_{\text{c}}$ ($M_{\text{b}}$) out of the $N_{\text{c}}$ ($N_{\text{b}}$) TLSs of the calorimeter (bath) are in their ground state. There are then several possible outcomes about the state of the whole system at the end of time interval $\Delta t$. 

(i) The qubit makes a jump to the ground state with the probability
\begin{equation} \label{e19}
\Delta p_\downarrow^{\text{q}} = |d|^2 g_{\text{c}}^2 \frac{M_{\text{c}}}{N_{\text{c}}} \Delta t.
\end{equation}
When this happens, the following changes take place: $M_{\text{c}}\rightarrow M_{\text{c}}-1$, $c \rightarrow 1$, $d\rightarrow 0$, and the energy (heat) released to the TLS calorimeter is $\Delta Q_{\text{c}}=+\hbar\omega_0$.

(ii) The qubit makes a jump to the excited state with the probability
\begin{equation} \label{e20}
\Delta p_\uparrow^{\text{q}} = |c|^2 g_{\text{c}}^2 \left(1-\frac{M_{\text{c}}}{N_{\text{c}}}\right) \Delta t.
\end{equation}
Then $M_{\text{c}}\rightarrow M_{\text{c}}+1$, $c \rightarrow 0$, $d\rightarrow 1$, and $\Delta Q_{\text{c}}=-\hbar\omega_0$. In both the processes (i) and (ii), the heat to the bath $\Delta Q_{\text{b}}$ vanishes.

If no jump occurs in the qubit within $\Delta t$, i.e., neither of the processes (i) or (ii) occurs, with probability $1- \Delta p_\downarrow^S -\Delta p_\uparrow^S$, the qubit evolves as
\begin{eqnarray} \label{e21}
&& \dot c = -\frac{i}{2\hbar}Fe^{i(\omega -\omega_0)t}d - g_{\text{c}}^2 \left(\frac{1}{2}- \frac{M_{\text{c}}}{N_{\text{c}}}\right)c|d|^2, \nonumber \\ && \dot d = -\frac{i}{2\hbar}Fe^{i(\omega_0-\omega) t}c + g_{\text{c}}^2 \left(\frac{1}{2}- \frac{M_{\text{c}}}{N_{\text{c}}}\right)|c|^2d. 
\end{eqnarray}

(iii) One of the calorimeter TLSs relaxes from the excited state to the ground state and one of the bath TLSs gets excited from the ground state to the excited state. The probability of this event is given by
\begin{equation} \label{e22}
\Delta p_\downarrow^{\text{c}} =  g_{\text{b}}^2 \left(1- \frac{M_{\text{c}}}{N_{\text{c}}}\right)\frac{M_{\text{b}}}{N_{\text{b}}} \Delta t.
\end{equation}
When this process takes place, it leads to the the following changes: $M_{\text{c}}\rightarrow M_{\text{c}}+1$, $M_{\text{b}}\rightarrow M_{\text{b}}-1$, and the heat released to the bath is $\Delta Q_{\text{b}}=+\hbar\omega_0$, and that to the calorimeter is $\Delta Q_{\text{c}} =-\hbar\omega_0$.

(iv) One of the calorimeter TLSs gets excited from the ground state to the excited state and one of the bath TLSs relaxes from the excited state to the ground state. The probability of this event is given by
\begin{equation} \label{e22a}
\Delta p_\uparrow^{\text{c}} =  g_{\text{b}}^2 \frac{M_{\text{c}}}{N_{\text{c}}}\left(1-\frac{M_{\text{b}}}{N_{\text{b}}}\right) \Delta t.
\end{equation}
This leads to $M_{\text{c}}\rightarrow M_{\text{c}}-1$, $M_{\text{b}}\rightarrow M_{\text{b}}+1$, $\Delta Q_{\text{b}}=-\hbar\omega_0$ and $\Delta Q_{\text{c}}=+\hbar\omega_0$. 

In case none of the processes (i)-(iv) occurs in this interval, with probability $1- \Delta p_\downarrow^{\text{q}} -\Delta p_\uparrow^{\text{q}}-\Delta p_\downarrow^{\text{c}}-\Delta p_\uparrow^{\text{c}}$,  $M_{\text{c}}$ and $M_{\text{b}}$ remain constant, and no energy is released into the calorimeter or the bath, $\Delta Q_{\text{c}} = \Delta Q_{\text{b}}=0$. As stated above, here the qubit obeys Eq.~\eqref{e21}.

\section{Distributions of heat and work} \label{distributions}

\begin{figure}[b]
    \begin{center}
    \includegraphics[width=.9\columnwidth]{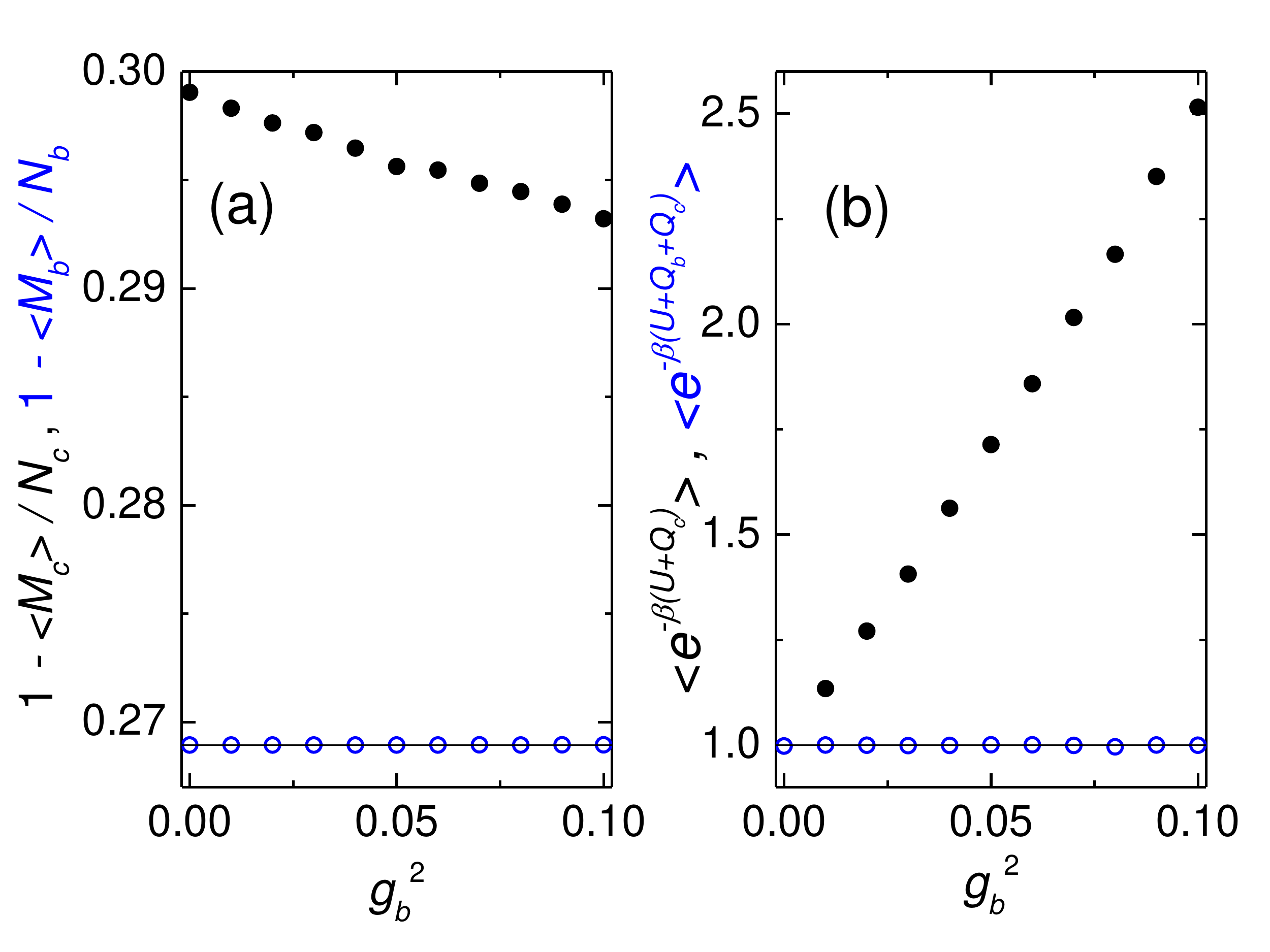}
    \end{center}
    \caption{\label{fig3} Results of a numerical simulation ($3\times 10^5$ repetitions for each data point) on a system of Fig. \ref{fig1} (c). The parameters are $N_{\text{c}}=10$, $N_{\text{b}}=10^4$, $\beta\hbar\omega_0=1$, $g_{\text{c}}^2 =0.1$, $F/2\hbar\omega_0=0.05$. The qubit is driven by a $\pi$-pulse. In (a) the average populations of the excited state, $1-\langle M_{\text{c}}\rangle/N_{\text{c}}$ (black) and $1-\langle M_{\text{b}}\rangle/N_{\text{c}}$ (blue), are shown as a function of the coupling $g_{\text{b}}^2$. The equilibrium population, $1/(1+e^{\beta\hbar\omega_0})$ is indicated by the horizontal solid line. The measurement of the populations $M_{\text{c}},M_{\text{c}}$ is done at the end of the $\pi$-pulse. (b) Test of the Jarzynski equality (indicated by the horizontal solid line) for the two expressions of work as given in Fig.~\ref{fig2}. Jarzynski equality is valid only for $W=U+Q_{\text{b}}+Q_{\text{c}}$.}
\end{figure}  
\begin{figure}[b]
    \begin{center}
    \includegraphics[width=.9\columnwidth]{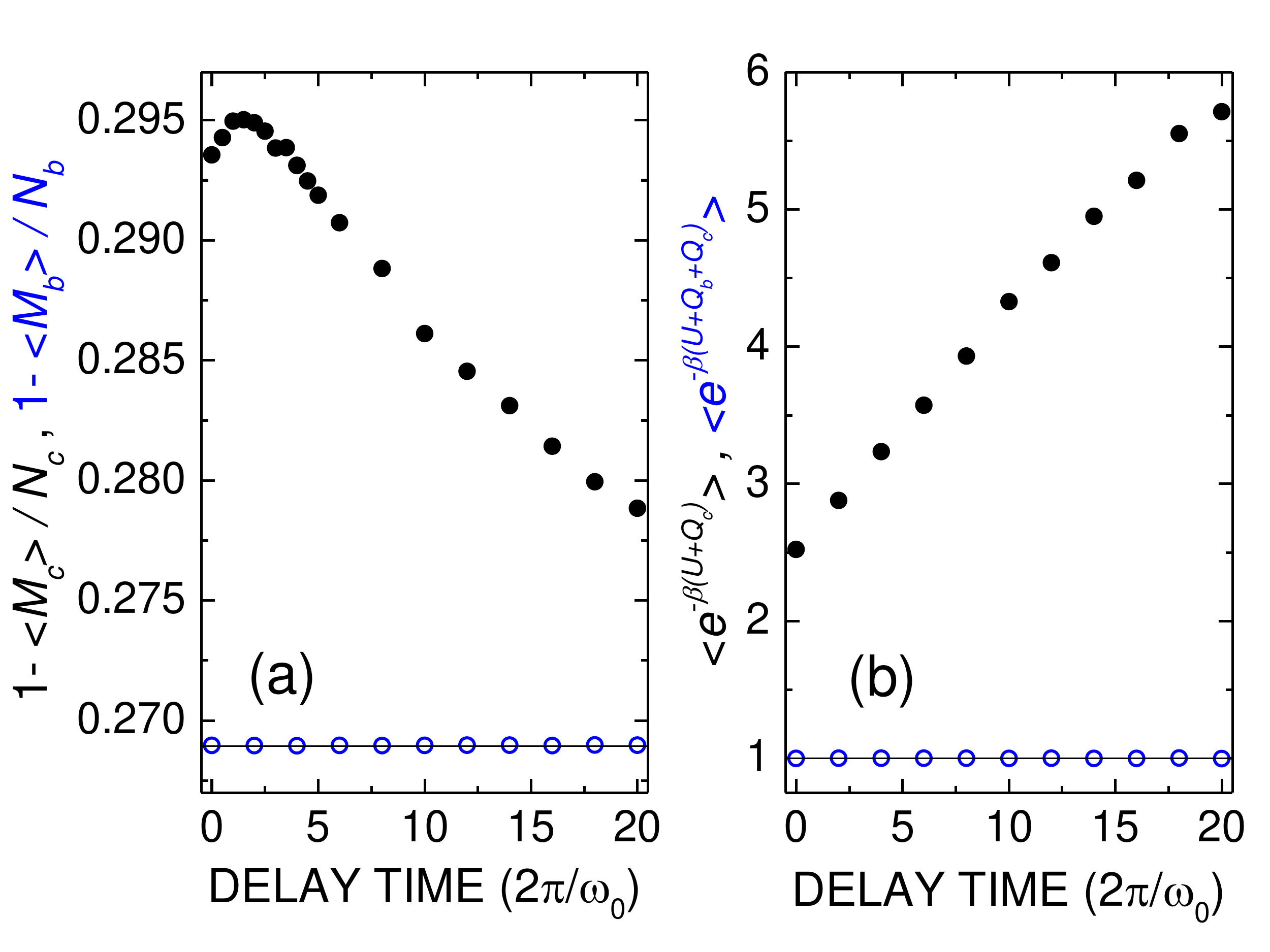}
    \end{center}
    \caption{\label{fig4} Results of a numerical simulation ($3\times 10^5$ repetitions for each data point) on a system of Fig. \ref{fig1} (c). The parameters are $N_{\text{c}}=10$, $N_{\text{b}}=10^4$, $\beta\hbar\omega_0=1$, $g_{\text{c}}^2 =g_{\text{b}}^2=0.1$, $F/2\hbar\omega_0=0.05$. The qubit is driven by a $\pi$-pulse. In (a) the average populations of the excited state, $1-\langle M_{\text{c}}\rangle /N_{\text{c}}$ (black) and $1-\langle M_{\text{b}}\rangle /N_{\text{b}}$ (blue), are shown as a function of the delay time after the $\pi$-pulse before the populations are measured. This delay time is given in units of $2\pi/\omega_0$. The equilibrium population, $1/(1+e^{\beta\hbar\omega_0})$ is indicated by the horizontal solid line. (b) Test of the Jarzynski equality (indicated by the horizontal solid line) for the two expressions of work as a function of the delay time. Jarzynski equality is valid only for $W=U+Q_{\text{b}}+Q_{\text{c}}$.}
\end{figure}  
Based on the procedure described in the previous Section, the statistics of heat $Q_{\text{c}}$, $Q_{\text{b}}$, and $Q_{\text{c}}+Q_{\text{b}}$, and work $W$ can then be analyzed in a given protocol. $Q_i$ is the sum over all the heat $\Delta Q_i$ deposited and extracted in the evolution. The apparent work $W=U+Q$ depends on what heat $Q$ can be observed in the measurement. Here $U$ is the change of the internal energy of the qubit, e.g., in a two-measurement protocol \cite{Kurchan2000}. For simplicity and in order to check the validity of obvious fluctuation relations, we assume that the TLSs are initialized such that each of them is in the ground state with probability $p_g=(1+e^{-\beta\hbar\omega_0})^{-1}$. The subsystems, in the present case the qubit and the TLS's in the calorimeter (and in the bath), are in their canonical states (with no coupling), see, e.g., \cite{andrieux2009}.
This is a usual situation in an amorphous system where the distribution of the TLS's level splitting is broad and, as a result, resonant TLSs are located far from each other~\cite{Black}.

Thus the probability of the calorimeter to be in a state with $M_{\text{c}}$ TLS's in the ground state is given by 
$$P_0(M_{\text{c}}) = \left(
\begin{array}{c}
N_{\text{c}} \\ M_{\text{c}}
\end{array}
\right)p_g^{M_{\text{c}}}(1-p_g)^{N_{\text{c}}-M_{\text{c}}},$$ and similarly for the bath by replacing indices ``$\text{c}$" by ``$\text{b}$".

Figure~\ref{fig2} presents numerical results with our model for the system of Fig.~\ref{fig1}~(c). Apparently the heat distributions do not differ much from the work distributions for a qubit in most regimes of interest, since the internal energy can have only two possible values $0$ or $\hbar\omega_0$. Therefore we focus on work distributions, as they can be readily assessed with regard of common fluctuation relations. Figure~\ref{fig2} demonstrates that the Crooks relation~\cite{Crooks99} (and thus the Jarzynski equality~\cite{Jarzynski97a,Jarzynski97b}) are valid within our model  always when the coupling $g_{\text{b}}^2$ to the big bath vanishes. For non-zero $g_{\text{b}}^2$, these relations fail, if one measures only the net heat into the calorimeter.

Figure \ref{fig3} (a) shows the dependence of the relative number of TLSs in the excited state at the end of the $\pi$-pulse in the calorimeter and in the big bath as functions of $g_{\text{b}}^2$. Panel (b) shows the test of the Jarzynski equality for the two definitions of work against $g_{\text{b}}^2$.
Note that a typical ratio $M_{\text{c}}/N_{\text{c}} \approx 0.7$ leads to the estimate $T_{\text{c}}/T \approx 1.18$ according to Eq.~\eqref{tc1}. Even if the calorimeter is substantially heated, it does not prevent the FRs from being valid if all the energy flows are properly taken into account.

Figure \ref{fig4} shows the same quantities as in Fig.~\ref{fig3} but now against the delay time between the driving $\pi$-pulse and the measurement of the populations. In (a) we see that right after the end of the driving pulse, heat is released from the qubit to the calorimeter (demonstrated by increase of the average excited state population in the calorimeter), and later on the calorimeter releases the excess heat to the bath. The driving source of the qubit performs the work. No work is done after the driving is over. This is consistent with the behaviour of the ``Jarzynski average" for $U+Q_{\text{b}}+Q_{\text{c}}$ in Fig. \ref{fig4}, but not for the quantity $U+Q_{\text{c}}$, which depends on this delay time and is not equal to unity in general. 

It is possible to prove analytically that the Crooks and Jarzynski fluctuation relations are satisfied. We show this in the Appendix for the case that the maximum number of jumps within each trajectory is one (single- and zero-jump processes). Here we refer to the system of Fig. \ref{fig1} (a). 

In conclusion, the model presented allows us to make the following observations. (i) A finite bath (calorimeter) is driven into non-equilibrium, if it is not coupled sufficiently strongly to the true bath. (ii) An isolated qubit and calorimeter set-up (Fig.~\ref{fig1}~(a)) satisfies Crooks and Jarzynski fluctuation relations, if the initial state is taken to be a factorized canonical configuration. (iii) If the calorimeter is further coupled to the true bath, the fluctuation relations fail if only the net energies into the system and calorimeter are taken into account. If the heat to the bath is included, the FRs are naturally valid again. (iv) Since all the heat of the calorimeter and bath are transported via the calorimeter, it is possible to assess and satisfy the FRs if one can detect the heat input from the system to the calorimeter. This is possible if one can detect the single quanta of relaxation and excitation events of the qubit by the calorimeter. Experimental progress on the corresponding electronic calorimeter coupled to a phonon bath has been reported recently in Refs.~\onlinecite{gasparinetti15,viisanen15,govenius15}. 

\acknowledgments
The work was supported by the Academy of Finland, contracts \# 272218 and 284594. YMG is thankful to the Aalto University for hospitality. We thank Tapio Ala-Nissila, Joakim Bergli, Antti Kupiainen, Paolo Muratore-Ginanneschi and Kay Schwieger for useful discussions.

 \appendix*

\begin{widetext}
\section{Analytic treatment of fluctuation relations} \label{app1}

In what follows we discuss the Crooks and Jarzynski equalities for trajectories up to a single jump. We assume that the qubit is driven resonantly in time $\tau$. In the absence of jumps, the amplitudes of the wavefunction $|\psi(\tau)\rangle = c(\tau)|g\rangle + d(\tau)|e\rangle$ evolve, up to the first order in $\Delta \Gamma = \Gamma_\downarrow - \Gamma_\uparrow$ as~\cite{hp2013}
\begin{eqnarray}
|c_g(\tau,t_i;M_{\text{c}})|^2 &=& 1 - |d_g(\tau,t_i;M_{\text{c}})|^2 = \cos^2\frac{ F (\tau-t_i)}{2\hbar} + \frac{\hbar
\Delta \Gamma (M_{\text{c}})}{F} \cos \frac{F (\tau-t_i)}{2\hbar} \sin^3 \frac{F (\tau-t_i)}{2\hbar}, \label{perag}\\
|c_e(\tau,t_i;M_{\text{c}}|^2 &= &1 - |d_e(\tau,t_i;M_{\text{c}})|^2 = \sin^2\frac{ F (\tau-t_i)}{2\hbar} + \frac{\hbar
\Delta \Gamma (M_{\text{c}})}{F} \cos \frac{F (\tau-t_i)}{2\hbar} \sin^3 \frac{F (\tau-t_i)}{2\hbar}
\label{perae},
\end{eqnarray}
where subscripts $g$ and $e$ refer to the evolution starting at $\tau=t_i$ in the ground or the excited state, respectively. As written explicitly, the rates $\Gamma_{\uparrow,\downarrow}(M_{\text{c}})$ are to be understood as those corresponding to the instantaneous state of the calorimeter. It is useful to define
\begin{eqnarray} \label{pi_eg}
&&\Pi_{g,e}(t_2,t_1;M_{\text{c}}) = \int _{t_1}^{t_2} d\tau [\Gamma_\uparrow (M_{\text{c}}) |c_{g,e}(\tau,t_1;M_{\text{c}})|^2 +
\Gamma_\downarrow (M_{\text{c}}) |d_{g,e}(\tau,t_1;M_{\text{c}})|^2],
\end{eqnarray}
which yield the probabilities in the form $e^{-\Pi_{g,e}(t_2,t_1;M_{\text{c}})}$ of not making a jump in the time interval $[t_1,t_2]$ starting in the ground or the excited state at $\tau=t_1$. 
\end{widetext}
Integrating Eq. \eqref{pi_eg} using Eqs. \eqref{perag} and \eqref{perae}, we obtain
\begin{equation} \label{perpi}
\Pi_{g,e}(\tau,0;M_{\text{c}})= \frac{\Gamma_\Sigma (M_{\text{c}}) \tau}{2 }\mp \frac{\hbar \Delta \Gamma (M_{\text{c}})}{2
F} \sin \frac{F \tau }{\hbar}, 
\end{equation}
again up to the linear order in $\Gamma$'s. Here $\Gamma_{\Sigma}(M_{\text{c}})=\Gamma_{\downarrow}(M_{\text{c}})+\Gamma_{\uparrow}(M_{\text{c}})$.
As an illustrative example we choose in what follows the duration of the drive $t$ to correspond to a $\pi$-pulse, by setting $F t/\hbar = \pi$. 

\subsection*{Crooks and Jarzynski relations} 
We will evaluate the following expressions up to the linear order in $\Gamma$s. In particular we calculate the ratios $P_i(-W)/P_i(W)$, where $W$ is the work in a realization, and $i=0,1$ refers to the number of jumps in a trajectory. 

The no-jump trajectories can yield work values $W=-\hbar\omega_0, 0, +\hbar\omega_0$ depending on the outcome of the first and second measurement. Therefore in this case we evaluate $P_0(-\hbar\omega_0)/P_0(+\hbar\omega_0)$. $W=+\hbar\omega_0$ for the process where the first measurement finds the system in the ground state and the second one in the excited state, with the probability
\begin{equation} \label{cr1}
P_0(+\hbar\omega_0)=\left \langle p_ge^{-\Pi_g(t,0;M_{\text{c}})}|d_g(t,0;M_{\text{c}})|^2 \right \rangle_{M_{\text{c}}}, 
\end{equation} 
and correspondingly
\begin{equation} \label{cr2}
P_0(-\hbar\omega_0)=\left \langle p_ee^{-\Pi_e(t,0;M_{\text{c}})}|c_e(t,0;M_{\text{c}})|^2\right \rangle_{M_{\text{c}}}. 
\end{equation} 
For the $\pi$-pulse $\Pi_g(t,0;M_{\text{c}})=\Pi_e(t,0;M_{\text{c}})=\Gamma_\Sigma(M_{\text{c}})\tau/2$ and $|d_g(t,0;M_{\text{c}})|^2 = |c_e(t,0;M_{\text{c}})|^2 =1$, which yield
\begin{equation} \label{cr3}
P_0(-\hbar\omega_0)/P_0(+\hbar\omega_0)=p_e/p_g = e^{-\beta\hbar\omega_0}, 
\end{equation} 
i.e., the Crooks relation for no-jump trajectories. 

The trajectories with one jump yield $W=-2\hbar\omega_0,-\hbar\omega_0,0,+\hbar\omega_0$ or $+2\hbar\omega_0$. For instance, $W=-2\hbar\omega_0$ arises for the realizations where the system starts in the excited state, makes one jump to the excited state and is found eventually in the ground state. The probability of such a process is given by 
\begin{eqnarray} \label{cr4}
&&P_1(-2\hbar\omega_0)=\left \langle p_e \int_0^t d\tau e^{-\Pi_e(\tau,0;M_{\text{c}})}\Gamma_\uparrow (M_{\text{c}})
|c_e(\tau,0;M_{\text{c}})|^2 \right.
\nonumber \\  &&  \left.\qquad \times
e^{-\Pi_e(t,\tau;M_{\text{c}}+1)}|c_e(t,\tau;M_{\text{c}}+1)|^2 \right \rangle_{M_{\text{c}}}.
\end{eqnarray} 
In this case, for $P_1(-2\hbar\omega_0)$ up to linear in $\Gamma$s, we may set $e^{-\Pi_i(t_1,t_2;M_{\text{c}})}=1$, and we may drop the $\Delta \Gamma$ dependence in the populations yielding $|c_e(t_1,t_2;M_{\text{c}})|^2 =\sin^2 (\pi(t_1-t_2)/2t)$, and 
\begin{eqnarray} \label{cr5}
&& P_1(-2\hbar\omega_0)= p_e \langle  \Gamma_\uparrow (M_{\text{c}})\rangle_{M_{\text{c}}} \int_0^t d\tau\sin^2 \left(\frac{\pi \tau}{2t}\right)
\nonumber \\ && \quad \times
 \sin^2\left (\frac{\pi(1-\tau/t)}{2}\right) = \frac{1}{8} \langle  p_e\Gamma_\uparrow (M_{\text{c}})\rangle_{M_{\text{c}}}t.
\end{eqnarray} 
Similarly,
\begin{equation} \label{cr6}
P_1(+2\hbar\omega_0)= \frac{1}{8} p_g \langle  \Gamma_\downarrow (M_{\text{c}})\rangle_{M_{\text{c}}}t.
\end{equation}
Then,
\begin{equation} \label{cr7}
\frac{P_1(-2\hbar\omega_0)}{P_1(+2\hbar\omega_0)}= \frac{p_e}{p_g}\frac{\langle  \Gamma_\uparrow (M_{\text{c}})\rangle_{M_{\text{c}}}}{\langle  \Gamma_\downarrow (M_{\text{c}})\rangle_{M_{\text{c}}}}.
\end{equation}
For equilibrium populations as here, $\langle  \Gamma_{\uparrow,\downarrow} (M_{\text{c}})\rangle_{M_{\text{c}}} = f(\pm \hbar\omega_0)$, and 
\begin{equation} \label{cr8}
P_1(-2\hbar\omega_0)/P_1(+2\hbar\omega_0)= e^{-2\hbar\omega_0},
\end{equation} 
as expected.

One-jump trajectories leading to $W=\pm \hbar\omega_0$ are those where the system starts and ends in the same state, but makes one jump to the ground (excited) state in between. Therefore, for example for $W=+\hbar\omega_0$ we have
\begin{widetext}
\begin{eqnarray} \label{cr9}
P_1(+\hbar\omega_0)&&=\left \langle p_g \int_0^t d\tau e^{-\Pi_g(\tau,0;M_{\text{c}})}\Gamma_\downarrow |d_g(\tau,0;M_{\text{c}})|^2 e^{-\Pi_g(t,\tau;M_{\text{c}}-1)}|c_g(t,\tau;M_{\text{c}}-1)|^2 \right.
\nonumber\\&& \qquad  \left.
+ p_e \int_0^t d\tau e^{-\Pi_e(\tau,0;M_{\text{c}})}\Gamma_\downarrow |d_e(\tau,0;M_{\text{c}})|^2 e^{-\Pi_g(t,\tau;M_{\text{c}}-1)}|d_g(t,\tau;M_{\text{c}}-1)|^2 \right \rangle_{M_{\text{c}}}.
\end{eqnarray}
\end{widetext}
With similar arguments as above, we find
\begin{equation} \label{cr10}
P_1(+\hbar\omega_0)= \frac{3}{8}  \langle  \Gamma_\downarrow (M_{\text{c}})\rangle_{M_{\text{c}}}t,
\end{equation}
and analogously
\begin{equation} \label{cr11}
P_1(-\hbar\omega_0)= \frac{3}{8}  \langle  \Gamma_\uparrow (M_{\text{c}})\rangle_{M_{\text{c}}}t,
\end{equation}
and thus
\begin{equation} \label{cr12}
P_1(-\hbar\omega_0)/P_1(+\hbar\omega_0)= e^{-\hbar\omega_0},
\end{equation} 
again as expected. We have thus shown that the Crooks relation is valid for processes linear in $\Gamma$s, i.e., zero and one-jump trajectories. Collecting all these contributions, we have
\begin{equation} \label{cr13}
P(-k\hbar\omega_0)/P(+k\hbar\omega_0)= e^{-k\hbar\omega_0},
\end{equation}
for $k=0,1,2$, and $P(W)\equiv P_0(W)+P_1(W)$.

Finally, Jarzynski equality is valid since it (always) follows from Crooks equality. For our case of discrete values of $W$, it is seen by
\begin{eqnarray} \label{je00}
\langle e^{-\beta W} \rangle& =&\sum_j P(W_j)e^{-\beta W_j} = \sum_{k=-2}^2 P(k\hbar\omega_0) e^{-k\hbar\omega_0} 
\nonumber \\
&=& \sum_{k=-2}^2 P(-k\hbar\omega_0)
 = \sum_j P(W_j) = 1.
\end{eqnarray}


\end{document}